\documentclass[a4paper]{jpconf}
\usepackage{graphicx}

\newcommand{\ttb}{t \bar t}

\begin{document}
\title{Missing Top Properties}

\author{J. A. Aguilar-Saavedra}

\address{Departamento de F\'{\i}sica Te\'orica y del Cosmos, Universidad de Granada, E-18071 Granada}

\ead{jaas@ugr.es}

\begin{abstract}
We discuss top polarisation observables at the Tevatron and the LHC with special attention to some that have not been measured and provide new, independent information about the top polarisation.
\end{abstract}

\section{Introduction}

In 2011, the measurement by the CDF Collaboration of an anomalous forward-backward (FB) asymmetry in $\ttb$ production at the Tevatron drew a great attention into top quark physics and, in particular, it fostered the theoretical studies of top quark properties (see~\cite{Aguilar-Saavedra:2014kpa} for a review and references). The deviations in the FB asymmetry are smaller in some of the latest measurements using the full Tevatron dataset. But, irrespectively of the reason behind these deviations---new physics, systematic bias or statistical fluctuations---the related developments are very interesting on their own. Here we will focus on top polarisation observables, which provide a good opportunity to detect elusive new physics in top pair and single top production. We will begin by setting the theoretical framework, and then proceed to discuss some polarisation observables for the Tevatron and for the Large Hadron Collider (LHC).

\section{Theoretical framework}
\label{sec:2}

The top quark is not a stable particle but it decays mainly into a $W$ boson and a $b$ quark, with a predicted average life of $5 \times 10^{-25}$ s. For a process in which a top quark is produced and subsequently decays into $Wb$, for example $pp \to t X \to W b X$ (with $X$ denoting possible additional particles), the amplitude contains an $s$-channel top propagator. Since the top quark width $\Gamma_t$ is small compared to its mass $m_t$, one can approximate the propagator as~\cite{Barger:1988jj}
\begin{equation}
\frac{\not\! p_t+m_t}{p_t^2-m_t^2 + i \Gamma_t m_t} \to \frac{\pi}{\Gamma_t m_t} \delta(p_t^2-m_t^2) \sum_\lambda u(p_t,\lambda) \bar u(p_t,\lambda) \,,
\label{ec:nw}
\end{equation}
in standard notation~\cite{Peskin:1995ev}, with $p_t$ the top quark momentum and $\lambda$ its helicity. Then, the amplitude is decomposed as $\mathcal{M} \propto \sum_\lambda A_\lambda \times B_\lambda$, a sum of production ($A_\lambda$) $\times$ decay ($B_\lambda$) factors, summed over the possible top helicities. Taking the modulus squared of the amplitude, the differential cross section can be written as $d\sigma \propto |\mathcal{M}|^2 \propto \sum_{\lambda,\lambda'} A_\lambda B_\lambda A_{\lambda'}^* B_{\lambda'}^*$. The product $A_{\lambda'}^* A_\lambda$ can be identified with the top spin density matrix, which for a spin-$1/2$ particle can be written in general as
\begin{equation}
\rho=\frac{1}{2}\left( \! \begin{array}{cc} 1+P_z & P_x-i P_y \\ P_x+i P_y & 1-P_z
\end{array} \! \right) \,,
\label{ec:rho}
\end{equation}
where $P_i = 2 \langle S_i \rangle$ is defined as twice the expected value of the three spin operators, $i=x,y,z$. In the present case we have chosen the $\hat z$ axis in the helicity direction, but the other two axes are not yet specified. $\vec P$ is a vector in three-dimensional space, satisfying $|\vec P| \leq 1$ in general, and $|\vec P| = 1$ if, and only if, the top quarks are produced in a pure spin state. Here is worth recalling that:
\begin{enumerate}
\item[(i)] The differential cross section {\it cannot} be decomposed into production $\times$ decay in general, since in general the off-diagonal elements in $\rho$ are non-vanishing. However, note that one can always choose a reference system $(x,y,z)$ where $\rho$ (a Hermitian matrix) is diagonal.
\item[(ii)] The off-diagonal elements involve the top quark polarisations $P_x$, $P_y$ in directions $\hat x$, $\hat y$ orthogonal to the $\hat z$ axis.
\item[(iii)] In the helicity basis these off-diagonal elements are given by the interference of helicity amplitudes with $\lambda \neq \lambda'$.
\item[(iv)] The integration over some variables~\cite{AguilarSaavedra:2012xe} can cancel the dependence on $P_{x,y}$ (hence also the effect of the quantum interference of helicity amplitudes) and in that case the ``classical'' production $\times$ decay interpretation is correct. 
\end{enumerate}

The top quark polarisation can be measured by using the double differential distribution of the charged lepton $\ell$ from the semileptonic $W$ decay $t \to W b \to \ell \nu b$ in the top quark rest frame (see for example~\cite{Godbole:2006tq}),
\begin{equation}
\frac{1}{\sigma}\frac{d\sigma}{d\!\cos \theta_\ell d\varphi_\ell} = \frac{1}{4\pi} \left(
1 + P_z \cos \theta_\ell + P_x \sin \theta_\ell \cos \varphi_\ell + P_y \sin \theta_\ell \sin \varphi_\ell
\right) \,,
\label{ec:dist2d}
\end{equation}
with $(\theta_\ell,\varphi_\ell)$ the polar coordinates of the charged lepton 3-momentum $\vec p_\ell$ in the $(x,y,z)$ reference frame. (Here we are ignoring possible $Wtb$ anomalous couplings in the top decay.)
After integration over $\varphi_\ell$, one recovers the well-known charged lepton polar angle distribution
\begin{equation}
\frac{1}{\sigma}\frac{d\sigma}{d\!\cos \theta_\ell} = \frac{1}{2} \left(
1 + P_i \cos \theta_\ell \right) \,,
\label{ec:dist1d}
\end{equation}
which allows to measure $P_z$. Notice that integration over $\varphi_\ell$ eliminates the dependence on $P_{x,y}$, so that for this distribution the production $\times$ decay interpretation of on-shell top quarks is adequate.

Alternatively to the full distribution (\ref{ec:dist2d}) one can use the one-dimensional polar angle distribution of the charged lepton with respect to the three axes~\cite{Baumgart:2013yra}, that is, Eq.~(\ref{ec:dist1d}) plus two more distributions obtained replacing $\theta_\ell \to \theta_\ell^i$ with $\theta_\ell^i$ the angle between $\vec p_\ell$ and the $i=x,y$ axis. These one-dimensional distributions can be experimentally simpler, and the information contained is the same: once the three polarisations are measured the spin density matrix (\ref{ec:rho}) is completely fixed.

In processes where several top (anti-)quarks are produced, the replacement (\ref{ec:nw}) can be done for each of them, which allows for the study of spin correlations in addition to the top polarisation considered here (see for example~\cite{Bernreuther:2004jv}).

\section{Polarisation at the Tevatron}
\label{sec:3}

\begin{figure}[htb]
\begin{center}
\includegraphics[width=7cm,clip=]{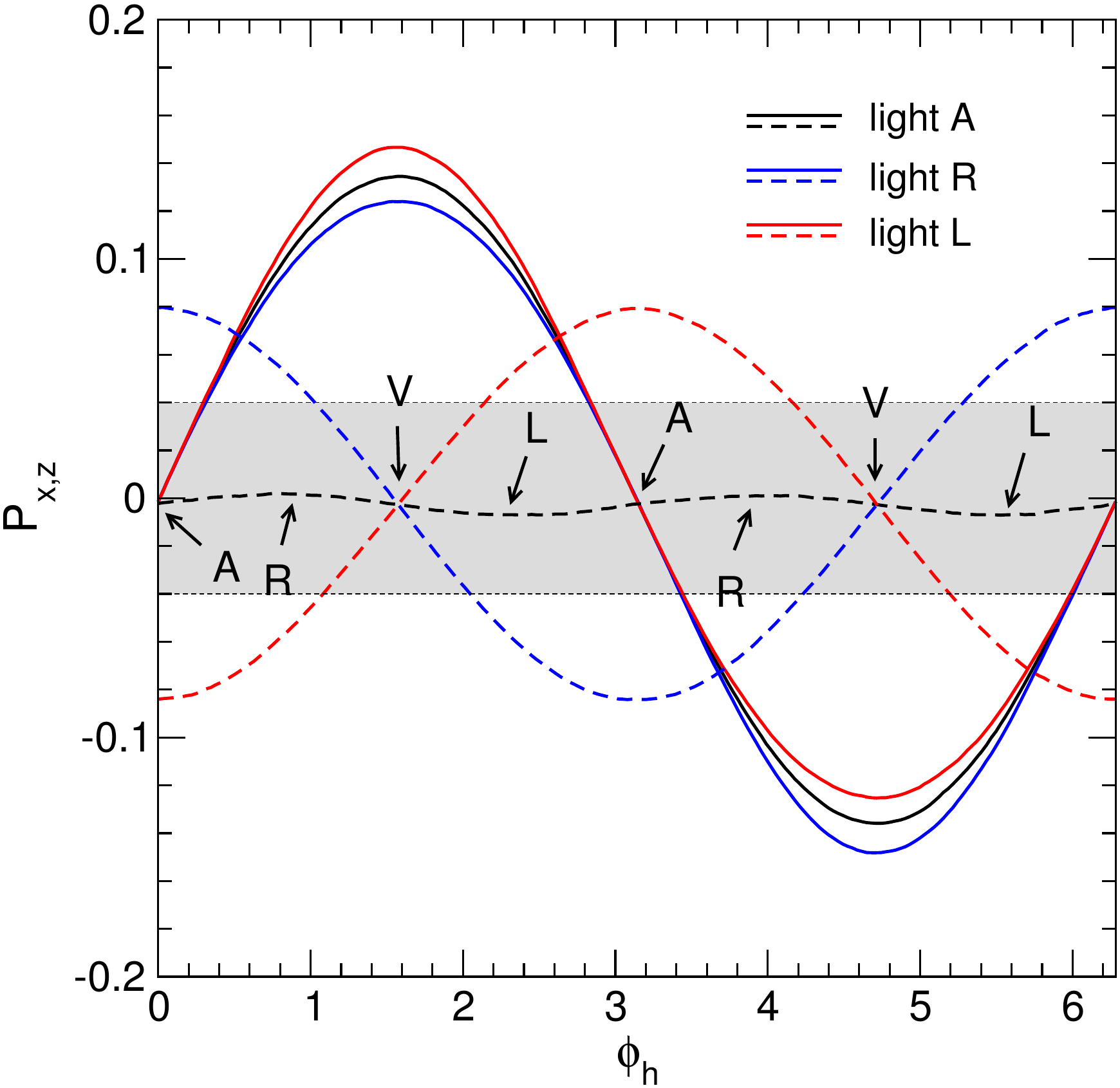}
\end{center}
\caption{Transverse (solid lines) and longitudinal (dashed lines) top polarisation as a function of the angle $\phi_h = \arg (g_V^t + i g_A^t)$ that parameterises the chirality of the top coupling. Points corresponding to top axial (A), vector (V), right-handed (R) and left-handed (L) couplings are indicated. The horizontal band is the statistical uncertainty for a typical sample of $\sim 2000$ events after selection~\cite{Aaltonen:2012it}. Adapted from~\cite{Aguilar-Saavedra:2014yea}. \label{fig:pol}}
\end{figure}

In order to measure the polarisation in the three axes we must specify our reference system, which at the Tevatron is rather easy because the $p \bar p$ collisions offer a privileged direction. The $\hat x$ axis can be taken in the plane spanned by the top quark momentum $\vec p_t$ and the proton momentum $\vec p_p$ in the centre-of-mass (CM) frame, and the $\hat y$ direction is then orthogonal to that plane. That is,
\begin{equation}
\hat z = \frac{\vec p_t}{|\vec p_t|} \,,\quad \hat y = \frac{\vec p_t \times \vec p_p}{|\vec p_t \times \vec p_p|} \,,\quad \hat x = \hat y \times \hat z \,.
\label{ec:xyz}
\end{equation}
The $\hat x$ and $\hat y$ directions are often called respectively ``transverse'' and ``normal''. The polarisations in these two directions are independent of the longitudinal polarisation $P_z$ usually considered. Let us consider $\ttb$ production and, as new physics benchmark, a colour octet with mass $M=250$ GeV and arbitrary coupling to the quarks,
\begin{equation}
\mathcal{L} = - \left[ \bar u \gamma^\mu {\textstyle \frac{\lambda^a}{2}} (g_V^u + \gamma_5 g_A^u) u 
+  \bar d \gamma^\mu {\textstyle \frac{\lambda^a}{2}} (g_V^d + \gamma_5 g_A^d) d 
+  \bar t \gamma^\mu {\textstyle \frac{\lambda^a}{2}} (g_V^t + \gamma_5 g_A^t) t 
\right] G_\mu^a  \,,
\end{equation}
with the condition $g_V^u-g_A^u = g_V^d-g_A^d$, {\it i.e.} the left-handed coupling to $u,d$ is the same. We plot in Fig.~\ref{fig:pol} the longitudinal and transverse polarisation for fixed $g_A^u [ (g_V^t)^2 + (g_A^t)^2 ]^{1/2} = 0.1$, three choices of the chirality of light quark couplings and varying chirality of the top quark coupling.
This plot not only demonstrates that $P_x$ and $P_z$ are independent and worth measuring, but also that $P_x$ can be (much) larger than the usually considered polarisation $P_z$. The normal polarisation $P_y$ can also be large~\cite{Baumgart:2013yra} but it requires non-trivial complex phases in the amplitude, which can originate from the $s$-channel propagator of the octet if it is very wide.

The same reference system in (\ref{ec:xyz}) can be used to other processes, {\it e.g.} single top production, but in this case it is unlikely that measurements are feasible at the Tevatron given the small size and purity of the signal.

\section{Polarisation at the LHC}
\label{sec:4}

\begin{figure}[t]
\begin{center}
\includegraphics[width=11cm,clip=]{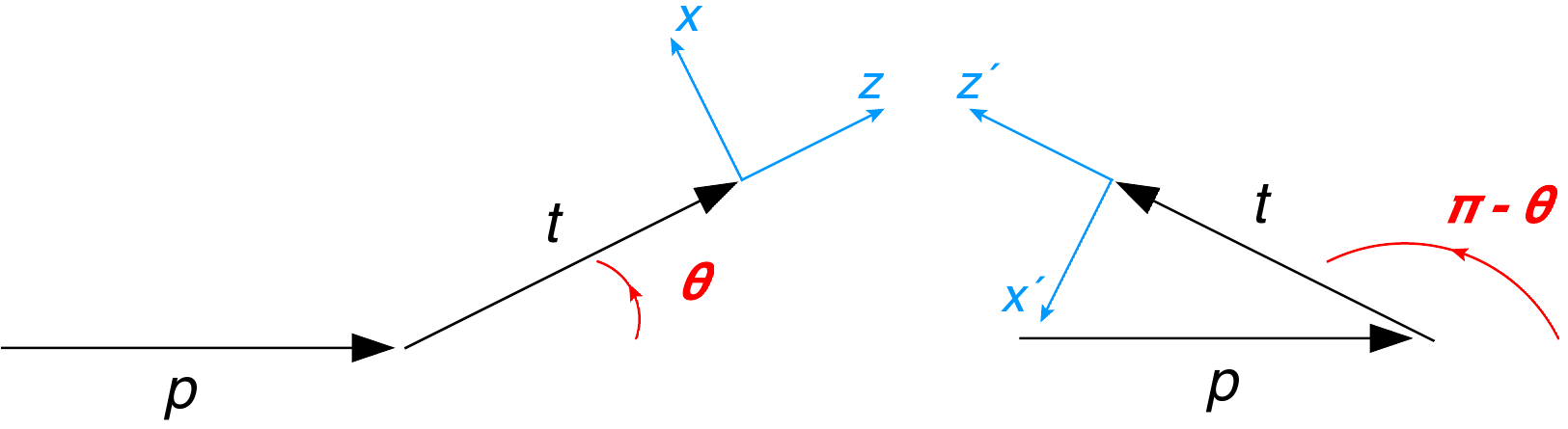}
\caption{Left: longitudinal and transverse directions for a top quark produced at an angle $\theta$ with respect to a fixed direction ($p$). Right: the same, for an angle $\pi - \theta$. The normal direction $\hat y = \hat z \times \hat x = \hat z' \times \hat x'$ is the same in both cases. \label{fig:xyz}}
\end{center}
\end{figure}

The definition of our reference system~(\ref{ec:xyz}) is not useful for the LHC because the two colliding hadrons are protons and there is no privileged direction in the initial state. Let us examine this in detail, considering $\ttb$ production as in the previous section and choosing the momentum of one of these protons $\vec p_p$ (fixed) to build our system as in (\ref{ec:xyz}). Then, since the initial quark in $q \bar q \to t \bar t$ can come from this proton or the other one, there are relations between the longitudinal, transverse and normal polarisations for scattering at opening angles $\theta$ and $\pi - \theta$, see Fig.~\ref{fig:xyz},
\begin{equation}
P_z(\theta) =   P_{z'}(\pi - \theta) \,,\quad
P_x(\theta) = - P_{x'}(\pi - \theta) \,,\quad 
P_y(\theta) = - P_{y'}(\pi - \theta) \,.
\label{ec:rel}
\end{equation}
Notice that we have labelled the reference system for opening angle $\pi-\theta$ as $(x',y',z')$ to remind the reader that the axis directions are different from the ones for angle $\theta$. The relations (\ref{ec:rel}) imply that after integration over $\theta$ the transverse and normal polarisations vanish.
Then, in order to have non-zero $P_{x,y}$, some alternatives have been proposed in the literature. One is to include a $\mathrm{sign} \cos \theta$ factor in the definition of observables~\cite{Bernreuther:1995cx}. This is equivalent to subtracting the polarisations in the forward minus the backward hemispheres. An alternative is to select the proton direction on a event by event basis. In $\ttb$ production this can be done by choosing the proton in the direction of the $z$ momentum of the $\ttb$ pair in the laboratory frame, that is, if $p^{z,\mathrm{lab}}_{\ttb} > 0$ we choose the proton with $p_p^{z,\mathrm{lab}} > 0$, and conversely~\cite{Baumgart:2013yra}. This amounts to guessing, event by event, which proton has provided the initial valence quark. For the benchmark studied in section~\ref{sec:3} the second option leads to larger integrated values of $P_x$. For illustration, we give in Table~\ref{tab:1} the transverse polarisations resulting with these two options for a CM energy of 8 TeV in the case of right-handed octet coupling to the light quarks and the top quark, and the longitudinal polarisation which is the same in both cases. We also include the value of $P_x$ that would result if the {\it true} proton providing the valence quark were always chosen.
These polarisations are smaller than current systematic uncertainties~\cite{Aad:2013ksa,Chatrchyan:2013wua} but they can be enhanced by imposing a lower cut on the velocity of the $\ttb$ pair $\beta_z^{t \bar t} = |p_t^z+p_{\bar t}^2| / (E_t + E_{\bar t})$.
\begin{table}[htb]
\caption{\label{tab:1}Transverse and longitudinal polarisation at the LHC for a selected benchmark point $\phi_h = \pi/4$ from Fig.~\ref{fig:pol}, and right-handed coupling to the light quarks.}
\begin{center}
\begin{tabular}{llll}
\br
$\mathrm{sign} \cos \theta$ & $P_x = 0.0021$ & & $P_z = 0.0126$\\
choose $p$                          & $P_x = 0.0106$ & (true $p$: $P_x = 0.0186$) & $P_z = 0.0126$\\
\br
\end{tabular}
\end{center}
\end{table}

\begin{figure}[htb]
\begin{center}
\includegraphics[width=8cm,clip=]{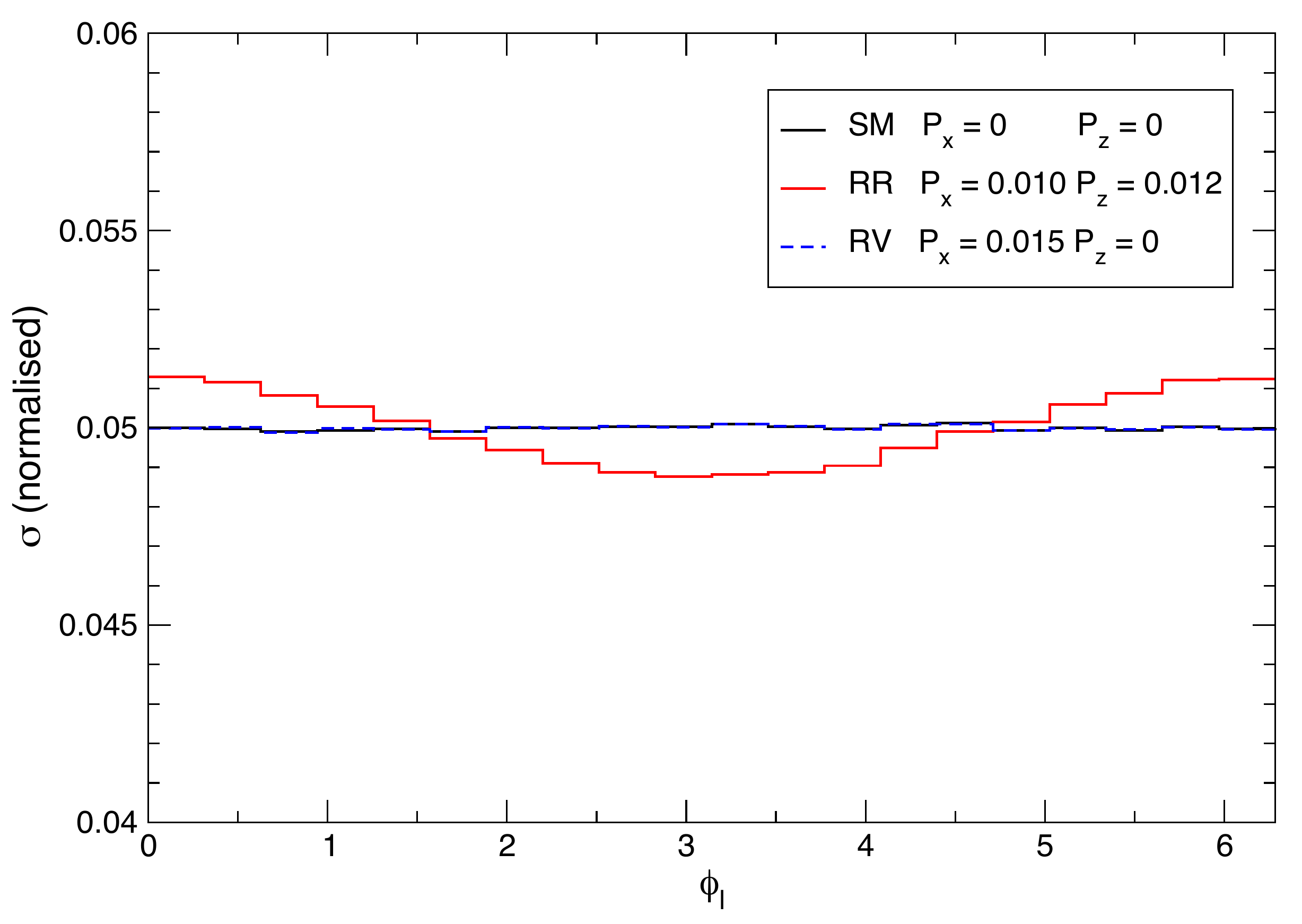}
\caption{Charged lepton azimuthal distribution in the $(u,v,w)$ reference system defined in the text, for three benchmark points. For clarity, the flat contribution from $gg \to \ttb$ is not included. \label{fig:az}}
\end{center}
\end{figure}

For completeness, let us comment that the azimuthal distribution~(\ref{ec:dist2d}) has also been proposed as an observable sensitive to the top polarisation~\cite{Godbole:2010kr}, but using a different reference system $(u,v,w)$ with $\hat w$ in the direction of a (fixed) proton $\vec p_p$, and $\hat v = \vec p_p \times \vec p_t / |\vec p_p \times \vec p_t|$, $\hat u = \hat v \times \hat w$. We plot this distribution in Fig.~(\ref{fig:az}) for three benchmark points: (i) the standard model (SM); (ii) the colour octet in Fig.~\ref{fig:pol} with $\phi_h = \pi/4$ and right-handed coupling to the light quarks (RR); (iii) the same but with vector coupling to the top quark, $\phi_h = \pi/2$ (RV). Note that in cases (i, iii) the longitudinal polarisation $P_z$ is zero, while the transverse polarisation $P_x$ is nonzero for (ii, iii). 

According to Eq.~(\ref{ec:dist2d}), the azimuthal distribution is sensitive to $P_u$ and $P_v$ of the ``alternative'' reference system $(u,v,w)$, so one might be led to consider that with this definition of the axes one is able to measure the transverse and normal polarisations (but in other directions $\hat u$, $\hat v$) even if the proton direction has been arbitrarily fixed. However, as we can clearly see in Fig.~\ref{fig:az}, the distribution is only sensitive to $P_z$ in the helicity direction, as it can be expected from the relations in~(\ref{ec:rel}). Therefore, the azimuthal distribution in this reference system does not offer any information independent from $P_z$.

In $t$-channel single top production one has the same ambiguity to choose between the two proton directions. However, one can guess which one provided the initial quark by the direction of the spectator jet in the laboratory frame~\cite{Aguilar-Saavedra:2014eqa}, which is very forward and following the initial quark direction in most cases. This choice gives a correct identification 95\% of the time for the production of top quarks, and 90\% for the production of top antiquarks, thus allowing for potentially clean measurements of the transverse and top polarisations in single top production.

\section*{Acknowledgements}
I thank the organisers of the Top 2014 conference for the excellent atmosphere that inspired many fruitful discussions. This work has been supported by the Spanish MICINN Project FPA2010-17915, MINECO Project FPA2013-47836-C3-2-P, and by the Junta de Andaluc\'{\i}a Projects FQM 101 and FQM 6552.


\section*{References}

\begin{thebibliography}{99}

\bibitem{Aguilar-Saavedra:2014kpa}
  Aguilar-Saavedra J A, Amidei D, Juste A and P\'erez-Victoria M 2014
  {\it Preprint} arXiv:1406.1798 [hep-ph]
  
\bibitem{Barger:1988jj}
  Barger V D, Ohnemus J and Phillips R J N 1989
  {\it Int.\ J.\ Mod.\ Phys.}\ A {\bf 4}  617

\bibitem{Peskin:1995ev}
  Peskin M E and Schroeder D V 1995
  {\it An Introduction to quantum field theory}
  (Reading, USA: Addison-Wesley)

\bibitem{AguilarSaavedra:2012xe}
  Aguilar-Saavedra J A and Herrero-Hahn R V 2013
  {\it Phys.\ Lett.}\ B {\bf 718} 983

\bibitem{Godbole:2006tq}
  Godbole R M, Rindani S D and Singh R K 2006
  {\it JHEP} {\bf 0612} 021

\bibitem{Baumgart:2013yra}
  Baumgart M and Tweedie B 2013
  {\it JHEP} {\bf 1308} 072

\bibitem{Bernreuther:2004jv}
  Bernreuther W, Brandenburg A, Si Z G and Uwer P 2004
  {\it Nucl.\ Phys.\ B} {\bf 690} 81

\bibitem{Aaltonen:2012it}
  Aaltonen T {\it et al.}  (CDF Collaboration) 2013
  {\it Phys.\ Rev.}\ D {\bf 87} 092002
  
\bibitem{Aguilar-Saavedra:2014yea}
  Aguilar-Saavedra J A 2014
  {\it Phys.\ Lett.}\ B {\bf 736} 132

\bibitem{Bernreuther:1995cx}
  Bernreuther W, Brandenburg A and Uwer P 1996
  {\it Phys.\ Lett.}\ B {\bf 368} 153

\bibitem{Aad:2013ksa}
  The ATLAS Collaboration 2013
  {\it Phys.\ Rev.\ Lett.}\  {\bf 111} 232002
  
\bibitem{Chatrchyan:2013wua}
  The CMS Collaboration 2014
  {\it Phys.\ Rev.\ Lett.}\  {\bf 112} 182001

\bibitem{Godbole:2010kr}
  Godbole R M, Rao K, Rindani S D and Singh R K 2010
  {\it JHEP} {\bf 1011} 144

\bibitem{Aguilar-Saavedra:2014eqa}
  Aguilar-Saavedra J A and dos Santos S A 2014
  {\it Phys.\ Rev.}\ D {\bf 89} 114009

\end{thebibliography}
\end{document}